\documentstyle[12pt,twoside,fleqn,npa,epsfig]{article}
\newcommand \beq{\begin{eqnarray}}
\newcommand \eeq{\end{eqnarray}}
\newcommand{\set}[2]{\newcommand{#1}{#2}}
\set{\pa}{\partial \over \partial\, }
\set{\ba}{\bar }
\set{\e}{\epsilon }                                        
\set{\intp}{\int{dp\over (2 \pi)^3}}
\set{\ppl}{(p+\frac q 2)}
\set{\pmi}{(p-\frac q 2)}
\set{\leftvector}{\stackrel{\leftarrow}{\partial }}
\set{\rightvector}{\stackrel{\rightarrow}{\partial }}

\title{Damping of giant resonances in asymmetric nuclear matter}

\author{K. Morawetz\address{Fachbereich Physik, Universit\"at Rostock,
                            18051 Rostock, Germany\\$^{\rm b}$Laboratorio 
                            Nazionale Del Sud, Via S. Sofia, 44-95123
                            Catania, Italy}\hbox{$^{\rm ,b}$},
        U. Fuhrmann\hbox{$^{\rm \theaddress}$} and
        R. Walke\hbox{$^{\rm \theaddress}$}\hbox{$^{\rm ,b}$} }

\begin{document}
\maketitle
\begin{abstract}
The giant collective modes in asymmetric nuclear matter
are investigated within a dynamic relaxation time approximation.
We derive a coupled dispersion relation and show that two sources
of coupling appear: (i) a coupling of isoscalar and isovector
modes due to different mean-fields acting and (ii) an explicit new
coupling in asymmetric matter due to collisional interaction. We
show that the latter one is responsible for a new mode arising
besides isovector and isoscalar modes. 
\end{abstract}

\section{Theoretical Preliminaries}

The treatment of collective modes in nuclear matter is documented in an 
enormous 
literature \cite{GRO96}. 
We want to investigate the collective excitations 
in asymmetric nuclear
matter \cite{FMW98,CLN97}. To this end we like to focus on the 
simplest microscopic theory which provides 
basic experimental features. This will be accomplished by a Fermi-gas 
model including dissipation.  
We consider the system consisting of a number
of different species (neutrons, protons, etc...) interacting with the 
own specie and with other ones. It is important to
consider the interaction between different
sorts of particles  if we want to include friction between different
streams of isospin components, etc. Especially the isospin current may
not be conserved by this way. Let us start with a set of coupled
quantum kinetic equations for the reduced density operator
$\rho_a$ for a specie $a$
\beq
\partial_t \rho_a(t) =i[\rho_a,{\cal E}_a+{\cal
U}(t)_a]-\sum\limits_b\int\limits_0^t
dt'{\rho_a(t')-\tilde \rho_b(t') \over \tau_{ab}(t-t')}\label{kin}
\eeq
where ${\cal E}={\cal P}^2/2m$ denotes the kinetic energy
operator and ${\cal U}$ the mean field operator and the external
field which is assumed
to be a nonlinear function of the density. We approximate the
collision integral by a non-markovian relaxation time. This
turned out to be necessary to reproduce damping of zero sound
\cite{AB92,MTM96}. It accounts for the fact that during a two-particle
collision a collective mode can couple to the scattering process.
Consequently, the dynamical
relaxation time represents the physical content of a hidden three-particle 
process and is an equivalent to the memory effects. Further we assumed
a relaxation with respect to a local equilibrium which we specify
by a small deviation of the chemical potential of specie $a$ 
determined by current conservation. Linearizing the kinetic 
equation (\ref{kin}) we obtain the matrix equation for the density
deviation $\ba n_b$ with respect to an external perturbation
$U^b_{\rm ext}$
\beq
&&\sum\limits_b \ba n_b \left\{
\delta_{ab}-{i \over \omega \tau_a+i} \left[\delta_{ab}
-{\tau_a \over \tau_b} C_{ab}\right] -\Pi_a(\omega+{i\over\tau_a}) 
                                               \alpha_{ab}\right\}
=\Pi_a(\omega+{i \over \tau_a}) U_{\rm ext}^a.\label{resp}
\eeq
The matrix $C_{ab}$ is given by
\beq
C_{ab}=\sum\limits_c \left\{ 1\over\tau \right \}_{ac}
{\Pi_c\Big[(\omega+{i\over \tau_a})\frac{m_a}{m_c}\Big]\over \Pi_c(0)}
\left\{ \frac{1}{\tau}\right\}^{-1}_{cb}\label{c1}
\eeq
with $\alpha_{ab}={\pa n_b} \ba U_a$ the
linearization of the mean field with respect to the deviation of
density from equilibrium value caused by the external
perturbation $U_{\rm ext}$. The partial polarization function of 
a specie $a$ is
\beq
\Pi_a(\omega)=2 \intp\frac{f_a\ppl-f_a\pmi}{\e_a\ppl-\e_a\pmi-\omega}.
\eeq
The factor 2 in front of the integral accounts for the spin
degeneracy according to $n_a=2 \intp f_a(p)$. 

Equation (\ref{resp}) represents the complete polarization of the
system, because $\ba n=\Pi \, U_{\rm ext}$. It represents a matrix
equation which is solved easily. The collective modes are given by
the zeros of the determinant of the matrix on the left hand side
of (\ref{resp}) because these are the poles of the polarization
matrix.
Since we
took into account relaxation processes between all species we are
able to cover current - current friction.
For a two component system, e.g. neutrons with density $n$ and
protons with density $p$, we write explicitly
\beq
&&(1-\Pi_n^{\rm M} \alpha_{nn})(1-\Pi_p^{\rm M}
\alpha_{pp})-(D_{np}+\Pi_n^{\rm M} \alpha_{np})(D_{pn}+\Pi_p^{\rm
M} \alpha_{pn})=0\label{dism}
\eeq
with the generalization of the Mermin polarization function \cite{M70} to
a multicomponent system
\beq
\Pi_a^{\rm M}={\Pi_a(\omega+{i\over \tau_a})\over
1-{i \over \omega\tau_a +i} (1- C_{aa})},
\label{mm}
\eeq
and the additional coupling due to asymmetry in the system
\beq
D_{np}={\tau_n\over \tau_p}{C_{np}\over C_{nn}-i \omega \tau_n}.
\eeq
The $D_{pn}$ are given by interchanging sort indices.
This term does not appear for symmetric matter. 
Therefore we call this term asymmetry coupling term further-on.

It is illustrative to see known results for symmetric nuclear matter.
This is performed for the case of equal relaxation times
$\tau_p=\tau_n=\tau$ and equal deviation from the
mean field $\alpha_1=\alpha_{nn}=\alpha_{pp}$ and
$\alpha_2=\alpha_{np}=\alpha_{pn}$. From (\ref{dism}) the known Mermin result 
of dispersion relation \cite{M70} is then obtained
\beq
1-(\alpha_{1}\pm\alpha_{2}) {\Pi(\omega+{i\over \tau})\over
1-{i \over \omega\tau +i} \left[1- {\Pi(\omega+{i\over
\tau}) \over \Pi(0)} \right]}=0\label{dis4}
\eeq
with the isovector mode $\alpha_{1}-\alpha_{2}$ and the
isoscalar mode $\alpha_{1}+\alpha_{2}$.

We have presented a general dispersion relation for the
multicomponent system including all known special cases. The
dispersion relation (\ref{dism}) is similar to the one derived
recently in \cite{CTL97} if we neglect the collisional coupling
$D_{np}$. Also a more general polarization function (\ref{mm}) 
is presented here including collisions within a conserving approximation
\cite{HPR93}. In the
following we will apply this expression for the damping of giant
dipole resonances for symmetric as well as for asymmetric nuclear
matter.

\section{Model for nuclear matter situation}

We connect the wave vector for giant resonances
according to the 
the Steinwedel-Jensen \cite{SJ68} model with the nuclear radius. 
If we assume that the density oscillation obeys a wave equation 
we get for the boundary condition that the radial velocity vanishes 
on the spherical surface with radius $R=1.13 A^{1/3}$ such that 
$j_l'(k R)\equiv 0$ with the spherical Bessel function of order 
$l=0,1...$ associated with the monopole, dipole... resonances.

Since monopole modes are compression modes we have no zero of first order 
for $l=0$.
For the dipole modes one has in first order $k=2.08/R$ which describes the 
giant isovector dipole resonance (IVGDR) while the ISGDR is a spurious mode 
in first order. This would just mean an unphysical oscillation of center of 
mass motion. However, in second order $k=5.94/R$ the isoscalar giant dipole 
resonance (ISGDR) has been observed recently \cite{D98} and citations therein. 
This can be considered as a density oscillation inside a sphere.
The resulting wave vectors have very low values compared with the Fermi wave
vector. This allows us to expand the Mermin polarization function
(\ref{mm}) with respect to small $q v_c/\omega$ ratios and $v_c$ the sound
velocity. We obtain
\beq
\Pi_a^{\rm M}(\omega)&=&\frac{n_a(\mu_a)}{m_a}{q^2 \over 
                       \omega (\omega+{i/\tau_a})},\label{entw}\\
n_a(\mu_a)&=& 2\,\lambda_a^{-3} f_{3/2}(z_a)\zeta_{corr} 
\eeq
where the thermal wave length is $\lambda_a^2=2 \pi \hbar^3/(m_a T)$,
$f_{3/2}$ the standard Fermi integral and $z_a={\rm e}^{\mu_a/T}$
the fugacity. 
The correction constant $\zeta_{corr}=1.22$ is 
introduced to fit the numerical solution of the dispersion relation 
with the approximative expansion (\ref{entw}).
With the help of this expansion the dispersion relation
(\ref{dism}) takes the form
\beq
0&=&\bigg[\omega\Big(\omega+\frac{i}{\tau_n}\Big)-c_{nn}^2 q^2\bigg]
    \bigg[\omega\Big(\omega+\frac{i}{\tau_p}\Big)-c_{pp}^2 q^2\bigg]\nonumber\\
&&\!\!\!-\left[c_{np}^2+i{\tilde c_{np}^2 \over ( \omega +{i/\tau_n})\tau_p}
\right]\left[c_{pn}^2+i{\tilde c_{pn}^2 \over ( \omega +{i/\tau_p}) 
\tau_n}\right] q^4\label{order}
\eeq
with the partial sound velocities $c$ and $\tilde c$
\beq
c_{ab}^2=\alpha_{ab}\frac{n_a(\mu_a)}{m_a}, \qquad 
\tilde c_{ab}^2=\frac{T}{m_a}\frac{\frac{f_{3/2}(z_a)}{f_{1/2}(z_a)}
                                  -\frac{f_{3/2}(z_b)}{f_{1/2}(z_b)}}
                 {\frac{\tau_{ab}}{\tau_{bb}}-\frac{\tau_{aa}}{\tau_{ba}}}.  
\eeq

The dynamic relaxation time has been derived 
via Sommerfeld expansion \cite{FMW98} as
\beq
{1\over \tau_{ab}(\omega)}&=&{1\over \tau_{ab}(0)} \left[1+\frac 3 4
\left ({\omega \over \pi T}\right )^2 \right]\label{mem}
\eeq
for $a,b$ neutrons or protons respectively. The markovian relaxation time 
was given in terms of the cross section $\sigma_{ab}$ between specie $a$ 
and $b$ as $\tau^{-1}_{ab}={4 m\over 3 \hbar^3}\sigma_{ab} T^2$.
The dispersion relation (\ref{dism}) takes then the form of 
a polynomial of tenth (sixth) order 
corresponding to the inclusion of memory (in)dependent relaxation times. 
While most of these solutions
will just lead to parasitary solutions (${\rm Re}\,\omega\le0$) we
will get two coupled modes, i.e. the isoscalar and isovector
mode. Furthermore a third mode appears at extreme asymmetries and/or strong 
collisional coupling.

The damping rate in classical approximation is given by the solution of 
(\ref{dis4})  with (\ref{entw}) (long wavelength) 
and reads $\gamma=1/(2\tau)$ .
We recognize that the FWHM is just twice the damping rate $\Gamma=2 \gamma$. 
This has been recently emphasized \cite{TKL97}.
It has to be stressed that the experimental data are accessible by this FWHM.

We will use as an illustrative example the following mean field
parameterization of Vautherin \cite{VB72,BV94}
\beq
U_a=t_0 \left [ (1+{x_0\over 2}) (n_n+n_p)-(x_0+\frac 1 2) n_a \right ]
+{t_3 \over 4} \left[(n_n+n_p)^2-n_a^2\right]
\eeq
with $a=n,p$ the density of neutrons or protons respectively.
The Coulomb interaction leads to an additional contribution for
the proton mean-field
\beq
U_p^{C}(q)={4 \pi e^2\over q^2} n_p(q).
\eeq
The here used model parameters reproduce the Weizs\"acker formula
\beq
{E\over A}= -a_1 +{a_2\over A^{1/3}}+{a_3 Z^2\over A^{4/3}} + a_4 \delta^2
\eeq
by the volume energy $a_1=15.68$ MeV, Coulomb energy $a_3=0.717$ MeV and 
the symmetry energy $a_4=28.1$ MeV with the asymmetry parameter
$\delta={n_n-n_p\over n_n+n_p}$.

\begin{figure}[t]
 \begin{minipage}[t]{7cm}
  \centerline{\epsfig{file=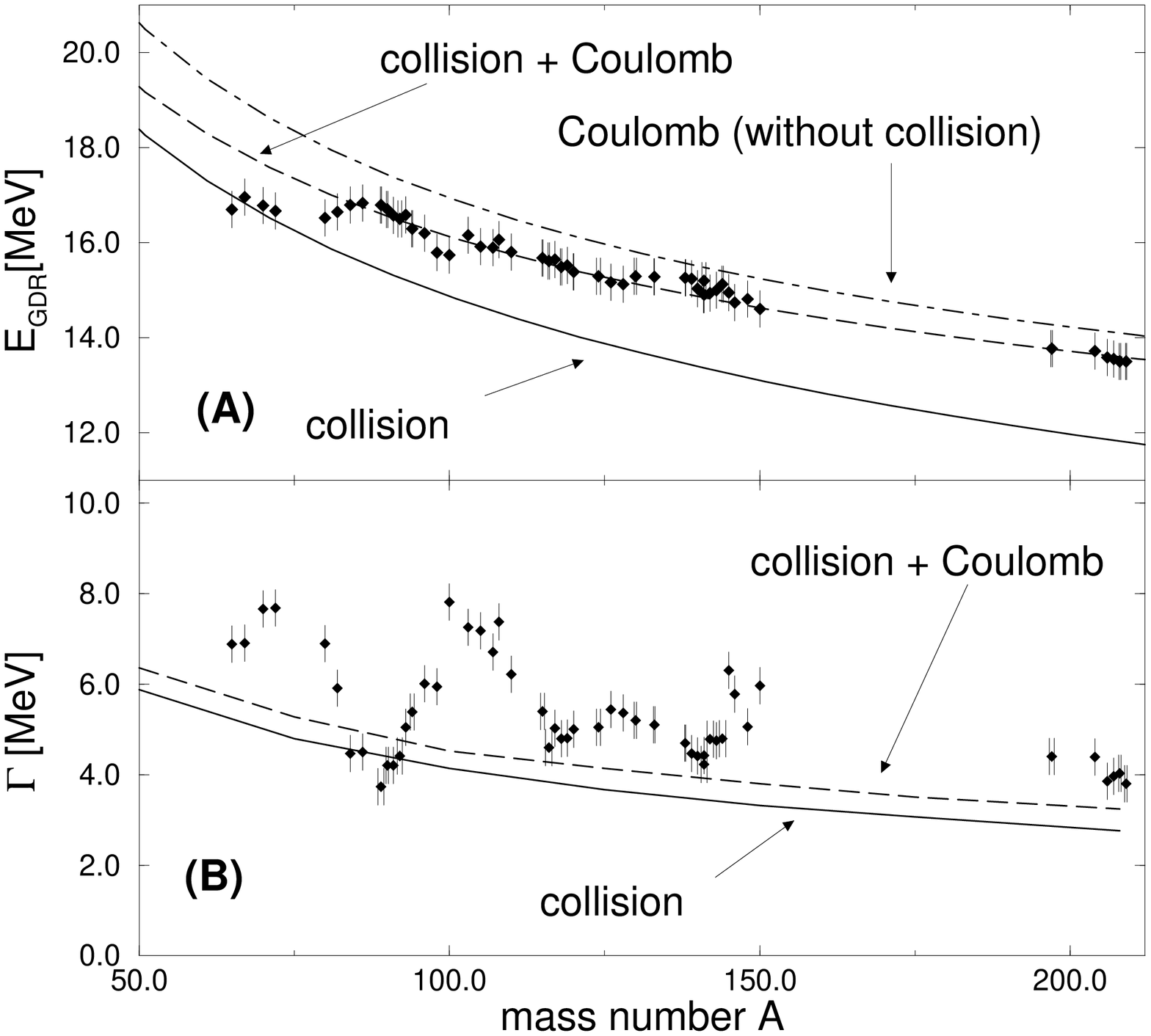,scale=.3,angle=-90}}
   \vspace{-.6cm}
   \begin{center}
    \parbox{6.9cm}{
     \caption{The experimental centroid energies {\bf (A)} 
              and damping rates {\bf (B)} [filled symbols] of the giant dipole
              resonances vs mass number (data from Ref. \protect\cite{BER88})
              together with different approximations at $T=0$. }
             \label{energie}    }
   \end{center}
 \end{minipage}
 \hspace{\fill}
 \begin{minipage}[t]{7cm}
  \centerline{\epsfig{file=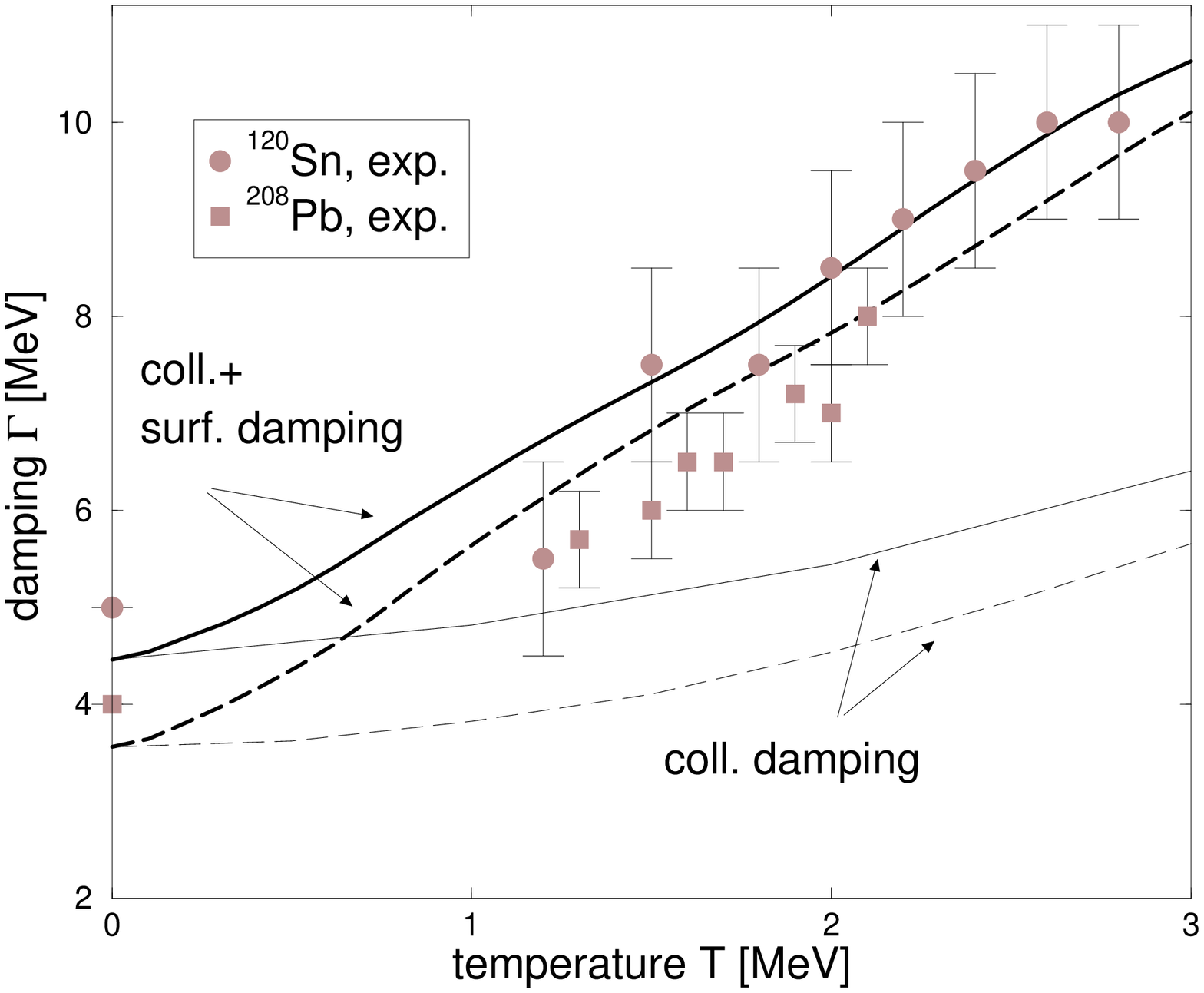,scale=.3,angle=-90}}
   \vspace{.05cm}
   \begin{center}
    \parbox{6.9cm}{
     \caption{The effective damping consisting of only collisional damping 
              compared with collisional and surface 
              damping [$^{120}$Sn (solid lines) and $^{208}$Pb (dashed lines)]
              together with the experimental data (filled symbols: Sn 
              from Ref. \protect\cite{RAP96} and Pb from 
              Ref. \protect\cite{RAM96}). } \label{pbt} }   
  \end{center}
 \end{minipage}
\end{figure}

First we plot the solution of the dispersion relation
(\ref{order}) for symmetric nuclear matter. In figure \ref{energie} 
we plotted the real and imaginary (FWHM) 
part of complex energy for different approximations with relaxation time 
(collisions) with and without Coulomb. In [figure \ref{energie} {\bf (A)}] we 
find that the
inclusion of Coulomb effects reproduces the experimental shape of 
the centroid energies at higher mass numbers (dot-dashed line). Taking
only collisions into account fails to describe higher mass numbers 
(solid line).
Considering Coulomb together with collisions (dashed line), the centroid
energies are  reduced towardes the data.   

The experimental values of the damping
rates are also presented versus mass number [figure \ref{energie} {\bf (B)}]. 
Without collisions we  would only have a vanishing Landau damping for
the infinite matter model \cite{FMW98}.
We see that the inclusion of Coulomb (dashed line) improves the situation.

The failure of the considered processes becomes more drastic if
we consider the temperature dependence of the GDR. In figure
\ref{pbt} we see that the temperature increase is too flat compared to
the experimental finding if we consider only collisional damping (thin lines). 
In  Ref. \cite{MVFLS98} we present a method to include also 
scattering with the nuclear surface. 
This improves the temperature dependence remarkably (thick lines).

Isoscalar giant dipole resonances (ISGDR) are observed recently 
in $^{208} Pb$ \cite{D98} and are given with a value 
\beq
E^{\rm ex}_{\rm ISGDR}=22.4\pm0.5\,{\rm MeV},
\qquad   \Gamma^{\rm ex}_{\rm ISGDR}=3.0\pm 0.5\,{\rm MeV}.
\eeq
They can be considered as a density oscillation inside a sphere. 
We obtain with the model parameters of IVGDR and a corresponding second 
order in $j_l'(k R)=0$
\beq
E_{\rm ISGDR}=21.0\,{\rm MeV},\qquad\Gamma_{\rm ISGDR}=3.4\,{\rm MeV}
\eeq
which is quite acceptable and is in the middle of different theoretical 
treatments (\cite{D98} and citations therein).

\section{New collective mode}

\begin{figure}
\begin{minipage}{7cm}
\centerline{\epsfig{file=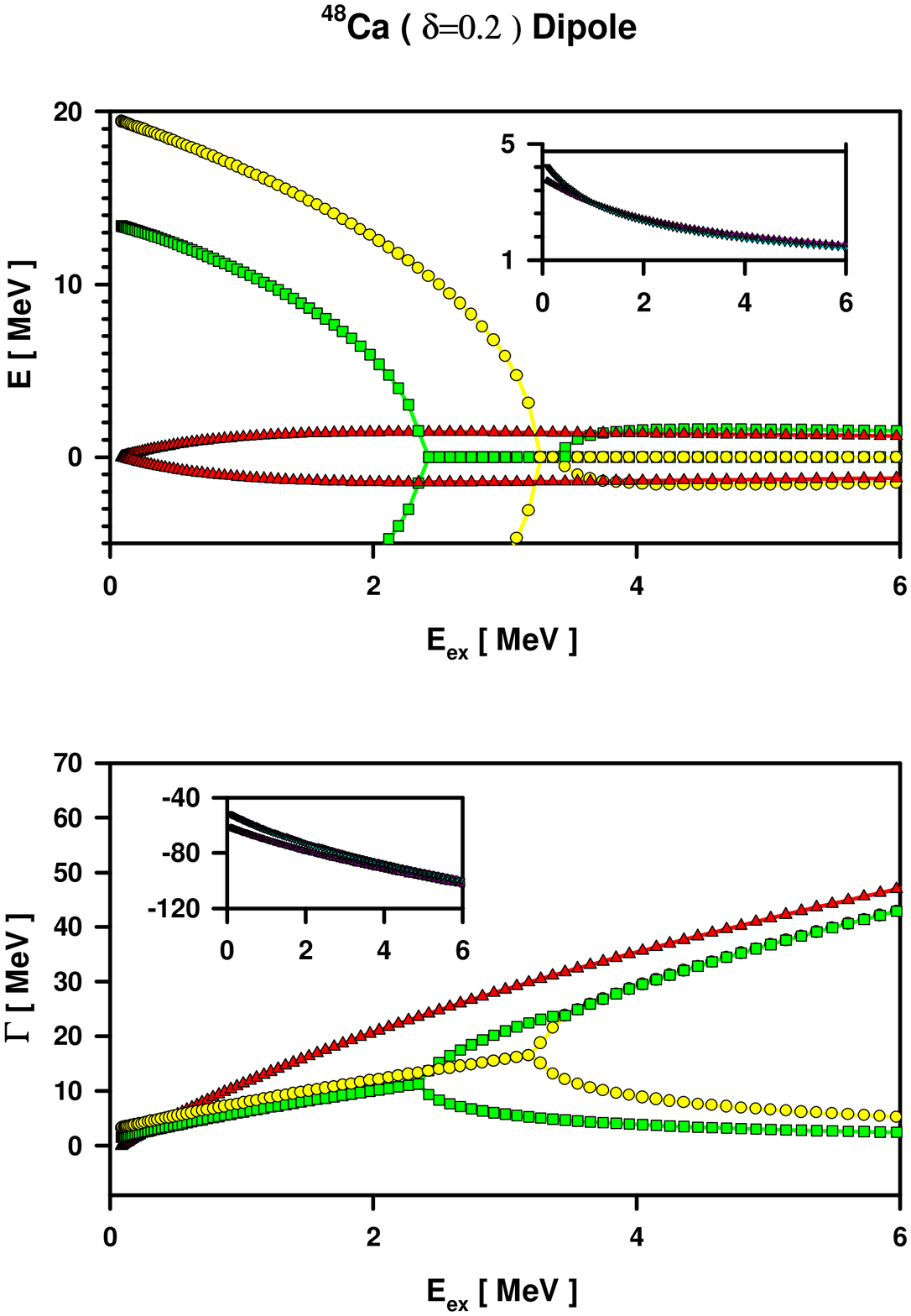,scale=.3,angle=0}}
\end{minipage}
\hspace{.1cm}
\begin{minipage}{7cm}
\centerline{\epsfig{file=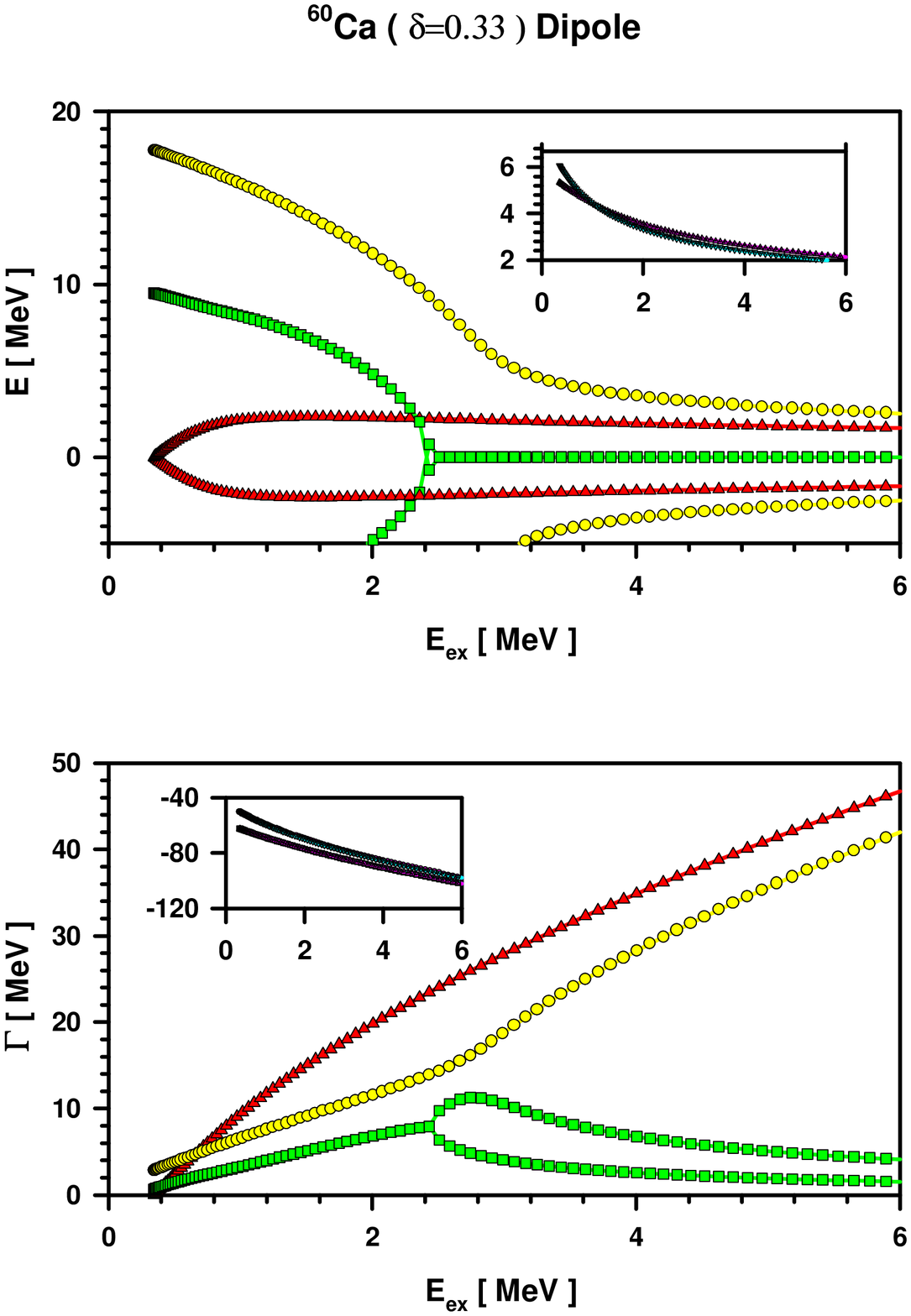,scale=.3,angle=0}}
\end{minipage}
\vspace{-1.2cm}
 \caption{The energy and damping of the IVGDR and ISGDR of $^{48}Ca$ (left) 
          and $^{60}Ca$ (right) vs excitation energy. 
          Besides the isovector modes (circles) and 
          the isoscalar  modes (squares) 
          a soft third mode appears (triangles).}
                      \label{d2820}
\end{figure}

Now we employ the same potential as in the last section but
assume different neutron and proton densities.
In figure \ref{d2820} we plot the isoscalar and isovector modes
versus temperature for $^{48}Ca$ with a small asymmetry $\delta=0.2$ as well
as for $^{60}Ca$ with a asymmetry $\delta=0.33$. 
The kinetic energy is
linked to a temperature within the Fermi liquid model via
Sommerfeld expansion 
\beq
{E\over A} &=&\frac 3 5 \epsilon_f \left[
{(1+\delta)^{5/3}+(1-\delta)^{5/3}\over 2} \right .+\left . {5\over 12}\pi^2 
\left({T\over T_f}\right)^2 {(1+\delta)^{1/3}+(1-\delta)^{1/3}\over 2} \right].
\label{fer}
\eeq
This connection between temperature and excitation energy is only valid for 
a continuous Fermi liquid model. For the small nuclei, the concept of 
temperature is questionable. Some improvement one can obtain by the 
definition of temperature via the logarithmic derivative of the density 
of states \cite{BM69}
\beq
T^{-1}=\frac{1}{\rho}{\partial \rho\over \partial E_{\rm ex}}
=-\frac  5 4 E_{\rm ex}^{-1}+\pi ({A\over 4 \epsilon_f E_{\rm ex}})^{1/2}
\eeq
which provides $E_{\rm ex}\approx \frac 1 4 (E/A)$ in comparison 
with (\ref{fer}) for small temperatures. We use this temperature to 
demonstrate possible collective bulk features in an exploratory sense. 
Of course, the surface energy and shell effects cannot be neglected for 
realistic calculations.

With increasing temperature the isovector and isoscalar energies decrease 
and vanish at a certain temperature. At these temperatures the damping 
becomes twofold because the damping of the spurious mode with negative 
energy becomes different from the physical mode. We can consider this 
behavior of damping as a phase transition of isospin demixture. 

At the same time a very soft mode appears due to collisional coupling which 
is only present in asymmetric matter \cite{MWF98}. This mode is more 
pronounced in the next figure \ref{d2820} for $^{60}Ca$ with an asymmetry 
of $\delta=0.33$.

We see that the isovector mode does not disappear 
but turns over into a flat decrease 
with increasing temperature. This behavior is coupled with the pronounced 
soft mode. In comparison with the more symmetric $^{48}Ca$ we see a different 
behavior of the damping where the isoscalar mode vanishes. 
Also the isovector mode appears 
unique and not two-folded. Now a clear transition of damping behavior for the 
isovector mode is recognizable 
which can be considered as a transition from zero to first sound damping.

\section{Summary}

To summarize we have observed that due to correlational coupling
there can exist a new mode which appears besides isovector and
isoscalar modes in asymmetric nuclear matter due to collisions. 
We suggest that
this mode may be possible to observe as a soft collective
excitation in asymmetric systems. The transition from zero sound damping to 
first sound damping behavior should become possible to observe for isovector 
modes since they do not vanish at this transition temperature like in 
symmetric matter.

The authors thank for the fruitful discussions
with M. DiToro, A. Larionov and V. Baran.
The hospitality of LNS-INFN Catania where part of this work has
been
finished, is gratefully acknowledged.
The work was supported by the Max-Planck Society Germany.


\begin{thebibliography}{10}

\bibitem{GRO96}
J.C. Bacelar and M.N. Harakeh and O. Scholten, {\it Proceedings of the Groningen
conference on giant resonances}, Nucl. Phys. A {\bf 599}, 1 (1996).
\bibitem{FMW98}
U. Fuhrmann, K. Morawetz, and R. Walke, Phys. Rev. C {\bf 58}, 1473 (1998).
\bibitem{CLN97}
F. Catara, E. G. Lanza and M. A. Nagarajan and A. Vitturi, 
Nucl. Phys. A {\bf 624},  449  (1996).
\bibitem{FMA98}
G. Fabbri and F. Matera, nucl-th/9805022.
\bibitem{AB92}
S. Ayik and D. Boilley, Phys. Lett. B {\bf 276},  263  (1992), errata ibd. 284
  (1992) 482.
\bibitem{MTM96}
K. Morawetz, M. DiToro, and L. M{\"u}nchow, Phys. Rev. C {\bf 54},  833
  (1996).
\bibitem{M70}
N.~D. Mermin, Phys. Rev. B {\bf 1},  2362  (1970).
\bibitem{CTL97}
M. Colonna, M. DiToro, and A.~B. Larionov, Phys. Rev. Lett.  (1997), sub.,
  Preprint LNS16-05-97.
\bibitem{HPR93}
H. Heiselberg, C.~J. Pethick, and D.~G. Ravenhall, Ann. Phys. {\bf 223},  37
  (1993).
\bibitem{SJ68}
H. Steinwedel and J. Jensen, Z. f. Naturforschung {\bf 5},  413  (1950).
\bibitem{D98}
B. Davis {\it et al.}, Phys. Rev. C  (1998), sub..
\bibitem{TKL97}
M. Di~Toro, V. Kolomietz, and A. Larionov,  in {\em proceedings of the Dubna
  conference on heavy ions} (unpublished, Dubna, 1997).
\bibitem{VB72}
D. Vautherin and D.~M. Brink, Phys. Rev. C {\bf 5},  626  (1972).
\bibitem{BV94}
F. Braghin and D. Vautherin, Phys. Lett. B {\bf 333},  289  (1994).
\bibitem{BER88}
S. Dietrich and B. Berman, Nucl. Data Tabl. {\bf 38},  199  (1988).
\bibitem{RAP96}
E. Ramkrishnan {\it et~al.}, Phys. Rev. Lett. {\bf 76},  2025  (1996).
\bibitem{RAM96}
E. Ramkrishnan {\it et~al.}, Nucl. Phys. A {\bf 549},  49  (1996).
\bibitem{MVFLS98}
K. Morawetz, M. Vogt, U. Fuhrmann, V. {\v S}pi{\v c}ka and P. Lipavsk\'y, 
Phys. Rev. C  (1998), sub..
\bibitem{BM69}
A. Bohr and B.~R. Mottelson, {\em Nuclear Structure} (W. A. Benjamin, Inc., New
  York, 1969).
\bibitem{MWF98}
K. Morawetz, R. Walke, U. Fuhrmann and M. DiToro,
Phys. Rev. C {\bf 57}, 2813 (1998).

\end{thebibliography}
\end{document}